\documentclass[a4paper]{article}

\usepackage{INTERSPEECH2022}
\usepackage{amsmath,graphicx}
\usepackage{amsfonts,amssymb}
\usepackage{booktabs}
\usepackage{cite}
\usepackage[font=small,labelfont=bf]{caption}
\usepackage{caption} 
\usepackage{xcolor}

\title{RCT: RANDOM CONSISTENCY TRAINING \\ FOR SEMI-SUPERVISED SOUND EVENT DETECTION}
\name{Nian Shao$^{1}$, Erfan Loweimi$^{2}$, Xiaofei Li$^{1}$\sthanks{ \ \ \ Corresponding author.}}
\address{
$^{1}$ Westlake University \& Westlake Institute for Advanced Study, Hangzhou, China \\
$^{2}$ Centre for Speech Technology Research (CSTR), University of Edinburgh, Edinburgh, UK}
\email{shaonian@westlake.edu.cn, e.loweimi@ed.ac.uk, lixiaofei@westlake.edu.cn}

\begin{document}

\maketitle
\begin{abstract}
  Sound event detection (SED), as a core module of acoustic environmental analysis, suffers from the problem of data deficiency. The integration of semi-supervised learning (SSL) largely mitigates such problem. This paper researches on several core modules of SSL, and introduces a random consistency training (RCT) strategy. First, a hard mixup data augmentation is proposed to account for the additive property of sounds. Second, a random augmentation scheme is applied to stochastically combine different types of data augmentation methods with high flexibility. Third, a self-consistency loss is proposed to be fused with the teacher-student model, aiming at stabilizing the training. Performance-wise, the proposed modules outperform their respective competitors, and as a whole the proposed SED strategies achieve 44.0\% and 67.1\% in terms of the $\text{PSDS}_1$ and $\text{PSDS}_2$ metrics proposed by the DCASE challenge, which notably outperforms other widely-used alternatives.
\end{abstract}

\section{Introduction}
\label{sec:intro}

Sound conveys a substantial amount of information about the environment. The skill of recognizing the surrounding environment is taken for granted by humans while it is a challenging task for machines \cite{virtanen2018computational}. Sound event detection (SED) aims to detect sound events within an audio stream by labeling the events as well as their corresponding occurrence timestamps. Taking advantage of deep neural networks, promising results have been obtained for SED \cite{turpault2019sound}. However, the high annotation cost poses obstacles on its further development. 

Two widely applied solutions for such data deficiency problem are data augmentation (DA) and semi-supervised learning (SSL). DA artificially enlarges the training dataset size in various forms including data warping, oversampling, etc. \cite{shorten2019survey}, while SSL leverages abundant unlabelled data to improve the model generalization capacity. For example, in computer vision (CV), different DA methods were proposed to transform the training images including rotation, cropping, etc. \cite{shorten2019survey}. To combine multiple augmentation methods, \textit{RandAugment} \cite{cubuk2020randaugment} presents a random strategy which arbitrarily selects one transformation in each training step. On the other hand, \textit{mixup} \cite{zhang2017mixup, tokozume2017learning} conducts a linear interpolation of two classes of data points to oversample the dataset, by which the mixed samples could push the decision boundaries into low-density regions. Essentially, DA regularizes the model by constraining the predicted labels to be invariant to any noise applied on the inputs. Such idea is known as \textit{consistency regularization} (CR) in SSL. When using CR for training, the model predictions are constrained to be invariant to any noise not only on the inputs \cite{verma2019interpolation} but also on the hidden states \cite{bachman2014learning, laine2016temporal}. However, there exists a potential risk known as \emph{confirmation bias} when the consistency loss is too heavily weighted in training \cite{tarvainen2017mean}. To alleviate such risk, \textit{MeanTeacher} \cite{tarvainen2017mean} applies a consistency constraint in the model parameter space, which holds an exponential moving average (EMA) of the training student model to generate pseudo labels for unlabelled data. Other techniques such as \textit{interpolation consistency training} (ICT) \cite{verma2019interpolation}, \textit{unsupervised data augmentation} (UDA) \cite{xie2019unsupervised} further combine mixup or RandAugmet with MeanTeacher and CR, obtaining state-of-the-art SSL models, respectively. 

The efforts on semi-supervised SED achieved promising results \cite{jiakai2018mean, Zheng2021, Nam2021, Gong2021, koh2021sound}, thanks to the Detection and Classification of Acoustic Scenes and Events (DCASE) challenge task  4\footnote{http://dcase.community/challenge2022}, which establishes a systematically organized semi-supervised dataset \cite{turpault2019sound} containing a reasonable amount of weakly-labelled (clip-level annotated) and unlabelled audio clips. The recent top-rank systems not only apply audio-specified DA methods including \textit{SpecAugment} \cite{park2019specaugment}, time shift \cite{koh2021sound}, pitch shift \cite{mcfee2015software}, etc., but also draw lessons from the SSL methods such as CR, MeanTeacher, etc. The top-rank system in 2018 DCASE challenge \cite{jiakai2018mean} introduces MeanTeacher into the semi-supervised SED, which becomes a role system for many variant systems \cite{Zheng2021, Nam2021, Gong2021, koh2021sound}. Similar to ICT \cite{verma2019interpolation}, shift consistency training (SCT) \cite{koh2021sound} proposes a way of unifying mixup, time shift and CR, obtaining a compatibly better semi-supervised SED model. Utilizing SCT and ICT, Zheng et al. \cite{Zheng2021} obtained the top rank in DCASE 2021 challenge task 4. Although multiple audio-specified DA methods \cite{park2019specaugment, koh2021sound, mcfee2015software} are leveraged in these models, the proper adaptation and effective combination of these SSL techniques are not scrutinised to be optimally applied in audio processing.

In this work, concerned with the proper usages of SSL techniques in audio processing, we propose a SSL strategy for SED, which consists of three novel modules:
\\
\indent i) \emph{Hard mixup}. Vanilla mixup \cite{zhang2017mixup} is trivially applied in many audio processing studies \cite{Kim2021, Tian2021, Lu2021}, while its adaptability for audio processing is not fully investigated. In vanilla mixup, the mixed label is a soft convex combination of the labels from original samples. Instead, Hard mixup preserves the hard label of the original samples, as the mixup of sound events yields concurrent sound events due to the additive property of audio signal. E.g.~The mixed label $[0.2, 0, 0.8]$ in vanilla mixup would be presented as $[1, 0, 1]$ in hard mixup. Hard mixup is the first algorithm which investigates the proper usage of the mixup method in the audio processing tasks.

\indent ii) \emph{RandomWarping}. Direct combination of various DA methods is not guaranteed to result in a performance gain, because of the complexity of finding an optimal set in a large hyperparameter searching space \cite{shorten2019survey}. Although many efforts have been made toward unifying various data augmentation schemes in CV \cite{cubuk2019autoaugment, cubuk2020randaugment}, they could not be trivially adapted for audio processing, since methods such as time shift \cite{koh2021sound} or pitch shift \cite{mcfee2015software} are specifically designed for audio. RandomWarping is among the first attempts toward unifying data augmentation methods for audio processing. We perturb each training sample with a randomly selected transformation, which allows taking advantage of different types of augmentation techniques in an unified way. It is a simple yet effective policy, while finding the optimal magnitude for each transformation can be challenging. We empirically find the optimal magnitude values for three data augmentation methods and achieve a consistent performance gain. \\
\indent iii) \emph{Self-consistency loss}. One of the challenges in SSL is designing the unsupervised loss for unlabelled data. In ICT \cite{verma2019interpolation} for mixup and SCT \cite{koh2021sound} for time shift, the MeanTeacher model and student model respectively process the original and augmented samples. In this paper, we propose an additional self-consistency loss to the MeanTeacher loss, which constraints the student model to give consistent predictions for original and augmented samples. Such self-consistency constraint always holds regardless of the correctness of the predictions, and thus would stabilize the training procedure. \\
\indent The combination of these three modules is referred to as \emph{random consistency training} (RCT), and the flowchart is shown in Fig.~\ref{figure: scheme}. With better adaptability for corresponding audio signals, each proposed module experimentally outperform its competing methods in the literature. As a whole, the proposed SSL strategy notably outperforms its counterparts, and achieves top performance on the DCASE 2021 challenge dataset. 

\section{The Proposed Method}
\label{sec:pagestyle}

SED is defined as a multi-class detection problem where the onset and offset timestamps of multiple sound events should be recognized from the input audio clips. We denote the time-frequency domain audio clips as $\mathbf{X}_i^{(l)} \in \mathbb{R}^{T \times K}$, where $T$ is the number of frames (same for all clips in experiment) and $K$ is the dimension of LogMel filterbank features. Three types of data annotations are used for training data, i.e.~weakly labelled, strongly labelled and unlabelled, which are indicated by superscript $\text{l}\in \{\text{w,s,u}\}$, respectively. $i$ denotes the index of data sample among a total of $N^{(\text{l})}$ data points of one annotation type. Let $C$ and $\mathbf{Y}_i^{(\text{l})}$ be the number of sound event classes and data labels, respectively. The weakly labelled and strongly labelled data have the clip-level and frame-level labels denoted by $\mathbf{Y}_i^{( \text{w} ) } \in \mathbb{R}^{C}$ and $\mathbf{Y}_i^{(\text{s})} \in \mathbb{R}^{T' \times C}$, respectively. Since the required time resolution for SED is much lower than the one of sound frames, pooling layers are applied in the CNN, resulting in a coarser time resolution $T'$ rather than $T$ for the predictions.

\begin{figure}[htb]
\begin{minipage}[b]{1.0\linewidth}
  \centering
  \centerline{\includegraphics[width=25em]{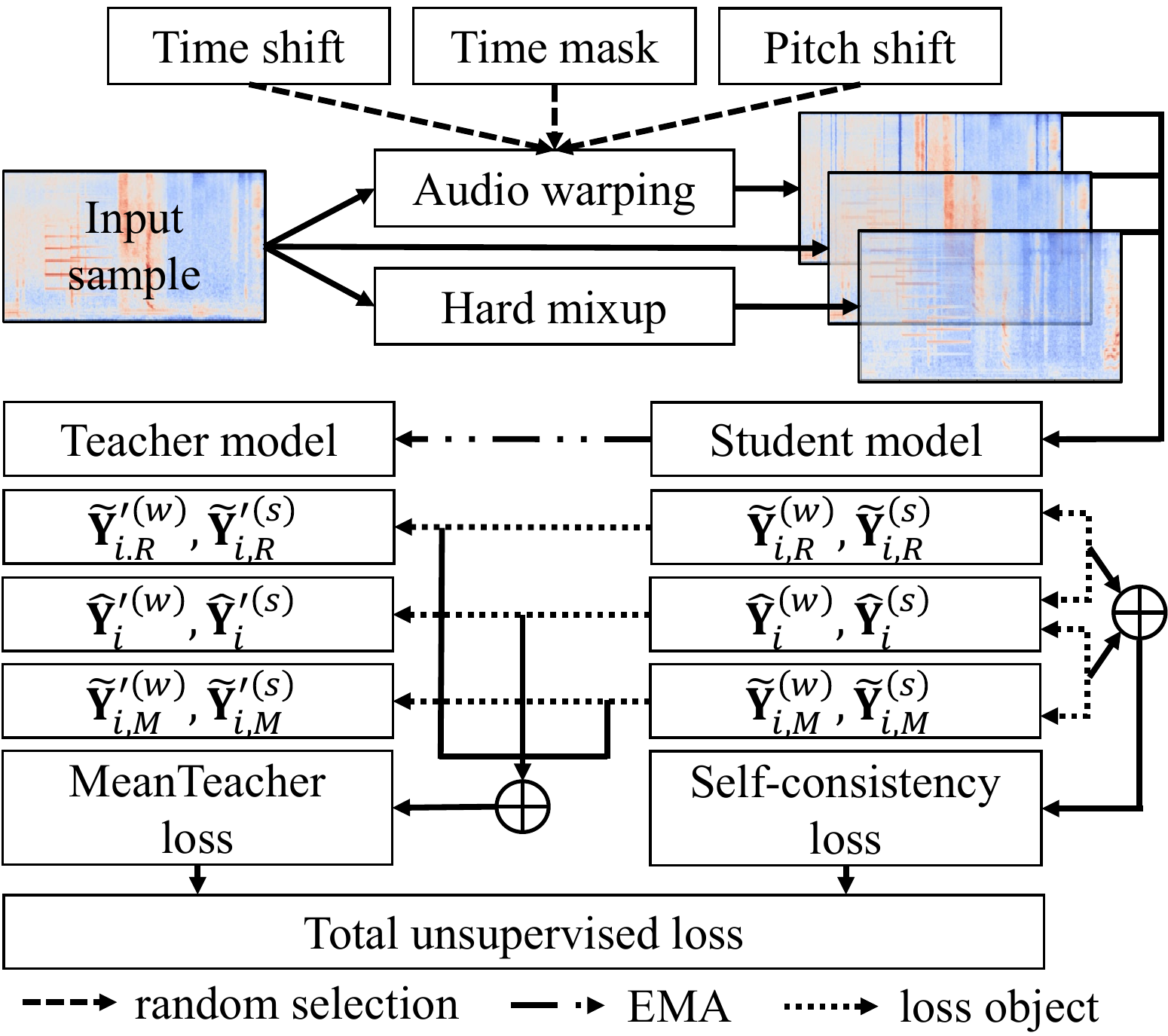}}
\end{minipage}
\vspace{-6.7mm}
\caption{Flowchart of RCT: both hard mixup and audio warping are first applied for data augmentation; MeanTeacher \cite{tarvainen2017mean} and self-consistency are used for SSL training. Subscripts $R$ and $M$ stand for RandomWarping and hard mixup, respectively.}
\label{figure: scheme}
\vspace{-7.3mm}
\end{figure}

The baseline model is a CRNN \cite{jiakai2018mean}, consisting of a 7-layer CNN with Context Gating layers \cite{Miech2017LearnablePW}, cascaded by a 2-layer bidirectional GRU. An attention module is added at the end to produce different levels of predictions \cite{jiakai2018mean} and MeanTeacher \cite{tarvainen2017mean} is employed for SSL. As shown in Fig.~\ref{figure: scheme}, a teacher model is obtained by an EMA of the student CRNN model to provide pseudo labels for unlabelled samples. The training loss is $\mathcal{L} = \mathcal{L}_\text{Supervised} + \mathcal{L}_\text{MeanTeacher}$, where $\mathcal{L}_\text{Supervised}$ is the cross-entropy loss for the labelled data, and $\mathcal{L}_\text{MeanTeacher}$ is the MeanTeacher mean square error (MSE) loss for the unlabelled data. 

\subsection{Random data augmentation for audio}
RandAugment \cite{cubuk2020randaugment} is proposed as an efficient way of combining different types of image transformations. In RCT, we take such idea to construct general audio warping methods for consistency regularization. Random data augmentation is accomplished by combining it with the proposed hard mixup, which means each training sample is augmented by hard mixup and random warping, respectively. As shown in Fig.~\ref{figure: scheme}, the batch size is thus tripled in each training step.

\textbf{Hard mixup}: The vanilla mixup \cite{zhang2017mixup} conducts an interpolation of two data points belonging to different classes, aiming to smooth the decision boundary. Such operation is feasible for images, whereas the interpolation of two audio clips produces a new clip due to the additive property of the sounds. As a result, combinations of multiple audio clips could be regarded as a realistic sample containing concurrent sound events, and should be recognized as an audio clip involving all sound events. Thus, the audio mixup is proposed to directly add multiple samples together, and the mixture is labelled with all the classes in all original samples. Since each sound clip reflect the realistic energies of all sound sources, the energy of the mixed sample is remained unchanged. Moreover, we noticed that combining more than two audio clips brings extra benefits. That is, it further condenses the distribution of sound events and can help the model toward better discriminating the sound events. Therefore, we randomly add two or three samples together in hard mixup.

\textbf{RandomWarping}: We use three audio warping methods in this work, and their warping magnitudes are set to $d\in \{1,2,\dots,9\}$. The only hyperparameter requires to be tuned is one unique $d_\text{max}$ for all audio warping methods. For each mini batch, one warping method is randomly chosen with a random magnitude $d$ uniformly distributed in $[1, d_\text{max}]$. This magnitude controls the audio warping intensities of the following methods: \\
\indent i) \textit{Time shift} \cite{koh2021sound} circularly shifts each audio clip along time axis with a duration of $1\times d$ seconds. \\
\indent ii) \textit{Time mask} \cite{park2019specaugment} randomly selects $5d$ intervals from the audio clip to be masked to 0. Since the shortest sound event lasts for 0.5 s, the length of each mask interval is set to 0.1 s. \\
\indent iii) \textit{Pitch shift} \cite{mcfee2015software} randomly raises or lowers the pitch of the audio clip by $1/2\times d$ semitones, where both pitches and formants are stretched. \\
The magnitude unit of each method ($1$ s in time shift, $0.1$ s in time mask, $1/2$ semitone in pitch shift) is empirically selected.

\subsection{Self-consistency training}
The MeanTeacher loss \cite{tarvainen2017mean} used in SED \cite{jiakai2018mean} has already shown a notable capacity in mining information from unlabelled data. To further leverage the unlabelled data, we propose to apply \emph{self-consistency} regularization in addition to the MeanTeacher loss. Let $\hat{\mathbf{Y}}_i^{(\text{w})}\in \mathbb{R}^C$ and $\tilde{\mathbf{Y}}_i^{(\text{w})}\in \mathbb{R}^C$ denote the weak (clip-level) predictions of original and augmented samples of the student model, respectively. Similarly let $\hat{\mathbf{Y}}_i^{(\text{s})}\in \mathbb{R}^{T' \times C}$ and $\tilde{\mathbf{Y}}_i^{(\text{s})}\in \mathbb{R}^{T' \times C}$ for the strong (frame-level) predictions. Self-consistency regulates the model by an extra MSE term added to the loss function
\begin{align}
    \mathcal{L}_{\text{SC}} = \ & r(step) \ \frac{1}{N^{(\text{w})}C} \sum_i^{N^{(\text{w})}} \|\mathcal{D}^{(\text{w})}_\text{aug}(\hat{\mathbf{Y}}_{i}^{(\text{w})}) - \Tilde{\mathbf{Y}}_{i}^{(\text{w})}\|_2^2 \notag \\
    + \ &
    r(step) \ \frac{1}{N^{(\text{s})}CT'}\sum_i^{N^{(\text{s})}} \|\mathcal{D}^{(\text{s})}_\text{aug}(\hat{\mathbf{Y}}^{(\text{s})}_{i}) - \Tilde{\mathbf{Y}}^{(\text{s})}_{i} \|_2^2,
\label{eq: self-consistency}
\end{align}
where $\|\cdot \|_2$ denotes the Euclidean norm for a vector/matrix, $r(step)$ is a ramp-up function varying along the training step. $\mathcal{D}_\text{aug}^{(\text{w})}$ and $\mathcal{D}_\text{aug}^{(\text{s})}$ denote transformations applied on the predictions of original samples, as the labels should be correspondingly changed for augmented samples. For pitch shift and time mask, there is no need to change the labels. Time shift should accordingly shift the strong labels along time axis. As for hard mixup, the mixed audio clip includes all the sound classes presented in the original audio clips. However, the labels for the combined sound classes cannot be trivially obtained by adding the predictions of original samples, since the summation of two soft-predictions/labels is meaningless. Instead, we define the label transformation for hard mixup as
\begin{align}
    \mathcal{D}^{(\text{l})}_{\text{mixup}}(\hat{\mathbf{Y}}^{(\text{l})}_{i}) = \vee_{i \in \mathcal{M}} \ \text{harden}(\hat{\mathbf{Y}}_{i}^{(\text{l})}),
\end{align}
where $\vee$ denotes element-wise OR operation, $\mathcal{M}$ is an arbitrary set consists of two or three data samples used in hard mixup, and $\text{harden}(\cdot)$ is an element-wise binary hardening function which ceils (or floors) the matrix elements to 1 (or 0) if the elements are larger than 0.95 (or smaller than 0.05). This transformation first hardens the predictions of original samples, from which the active/inactive sound classes are combined. 

The prediction of original and augmented samples ($\hat{\mathbf{Y}}_i$ and $\tilde{\mathbf{Y}}_i$) are both used for gradient updating, which means they are considered as the pseudo labels for each other in training. This is also one reason of choosing symmetric MSE function as an extra component in the loss function. The total loss $\mathcal{L}$ used for training the CRNN model will be
\begin{align}
    \mathcal{L} = \mathcal{L}_{\text{Supervised}} + \mathcal{L}_{\text{MeanTeacher}} + \mathcal{L}_{\text{SC}}, 
\end{align}
where $\mathcal{L}_{\text{MeanTeacher}}$ is the average of MeanTeacher MSE losses for the original sample and two augmented samples (hard mixup and random warping), as shown in Fig.~\ref{figure: scheme}. The unsupervised loss is composed of the MeanTeacher loss and self-consistency regularization. Such self-consistency constraint between the original and augmented samples always holds regardless of the correctness of the predictions.  This is different from ICT  \cite{verma2019interpolation} that merges the MeanTeacher loss and the consistency loss as one single loss. In ICT, the MeanTeacher loss is set as the MSE loss between the prediction of augmented and original samples, where the predictions are obtained by the teacher model and student model, respectively. Such pseudo labels highly rely on the correctness of the MeanTeacher predictions, while incorrect pseudo labels may mislead the student model and consequently reduce the training efficiency.

\begin{figure}[htb]
\begin{minipage}[b]{1.0\linewidth}
  \centering
  \centerline{\includegraphics[width=0.95\textwidth]{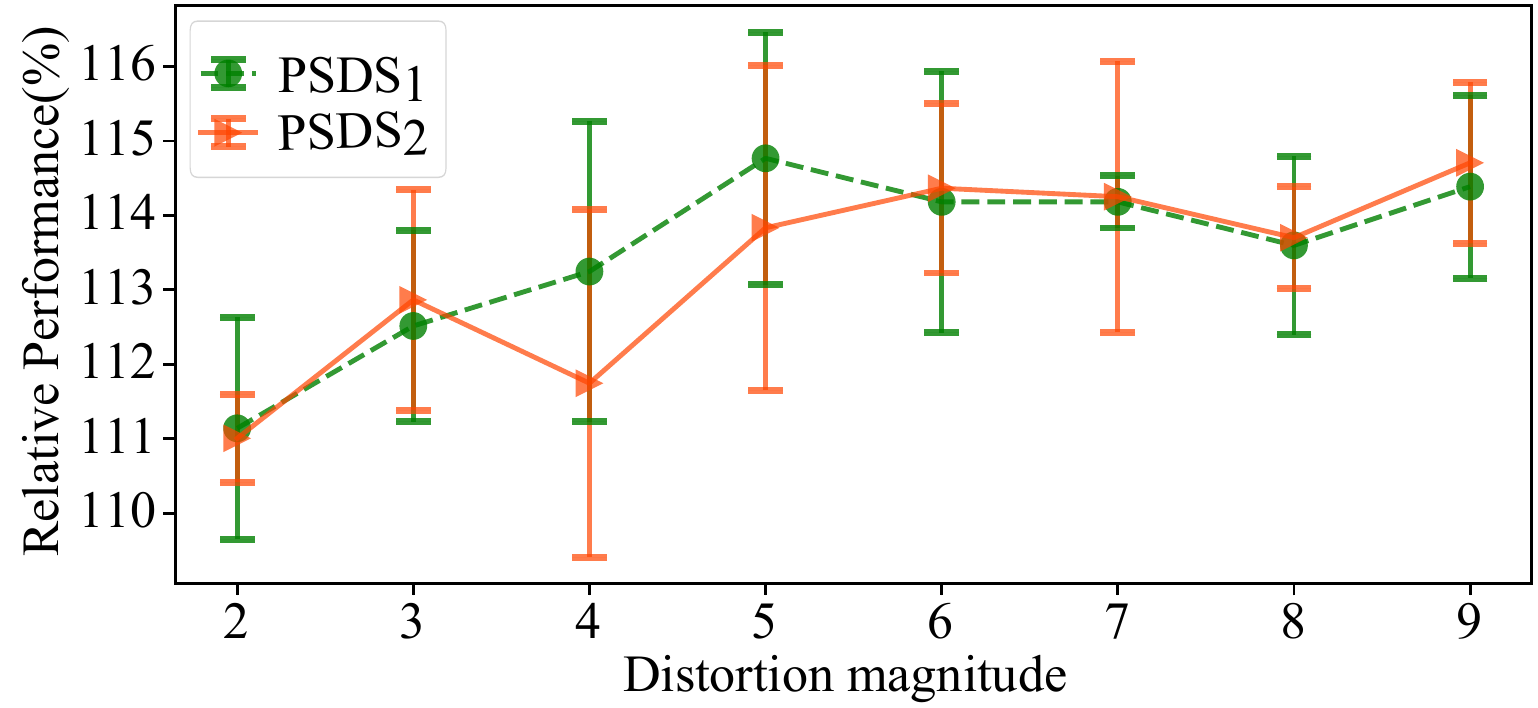}}
\end{minipage}
\caption{The relative performance gain as a function of maximum transformation magnitude ($d_{max}$). The transformation magnitude $d \sim U[1,d_\textit{max}]$. Relative performance gain is computed using the baseline performance ($\text{PSDS}_1=34.74\%$, $\text{PSDS}_2=53.66\%$). The markers and vertical lines represent the mean and standard deviation computed using three trials.}
\vspace{-1mm}
\label{figure: rank}
\end{figure}

\section{Experimental Results and Discussion}
\label{sec:typestyle}
Our codes for this work have been released on our website \footnote{https://github.com/Audio-WestlakeU/RCT}. We use the baseline model \cite{jiakai2018mean} on DCASE 2021 Task 4 dataset\footnote{http://dcase.community/challenge2021/task-sound-event-detection-and-separation-in-domestic-environments} to test the performance of the proposed method. The dataset consists of 1578 weakly-labelled, 10000 synthesized strongly-labelled and 14412 unlabelled audio clips. Each 10-second audio clip is resampled to 16 kHz and frame blocked with a length of 128 ms (2048 samples) and a hop length of 16 ms (256 samples). After 2048-point fast Fourier transform, 128-dimensional LogMel features are extracted for each frame, converting the 10-second audio clip into a $626 \times 128$ spectrogram. All samples are normalized to $[-1, 1]$ before training.

The batch size is set to 48, consisting of 12 weakly-labelled, 12 strongly-labelled and 24 unlabelled data points. The learning rate ramps up to $10^{-3}$ until epoch 50 and is scheduled by Adam optimizer \cite{kingma2015adam} until the end of training, i.e.~200 epochs. The weight for MeanTeacher and self-consistency losses, $r(step)$, linearly ramps up from 0 to 2 at epoch 50 and is then kept unchanged. The system performance is evaluated through polyphonic sound detection scores (PSDSs) \cite{mesaros2016metrics} according to the DCASE 2021 challenge guidelines. The metrics takes both response speed ($\text{PSDS}_1$) and cross-trigger performance ($\text{PSDS}_2$) into account; the larger the better for both metrics.

\subsection{Ablation study on SED}

\begin{table}[t!]
    \centering
    \caption{Ablation study for RCT. Different modules are added step by step and each score is obtained by averaging three trials.}
    \begin{tabular}{lcccc}
        \toprule
        \textbf{Model}  & \textbf{$\text{PSDS}_1$} (\%)  & \textbf{$\text{PSDS}_2$} (\%) \\
        \midrule
        Baseline                                  & 34.7 & 53.7 \\
            + Vanilla mixup \cite{zhang2017mixup} & 34.9 & 57.9 \\
            + Hard mixup                          & 36.4 & 57.4 \\
            \ \ \ + RandomWarping                 & 38.1 & 58.5 \\
            \ \ \ \ \ + ICT consistency\cite{verma2019interpolation} & 38.0 & 59.2 \\
            \ \ \ \ \ + Self-consistency          & 40.1 & 61.4 \\
        \bottomrule
    \end{tabular}
    \label{table: ablation_study}
    \vspace{-2mm}
\end{table}

\begin{figure}[htb]
\begin{minipage}[b]{1.0\linewidth}
  \centering
  \centerline{\includegraphics[width=\textwidth]{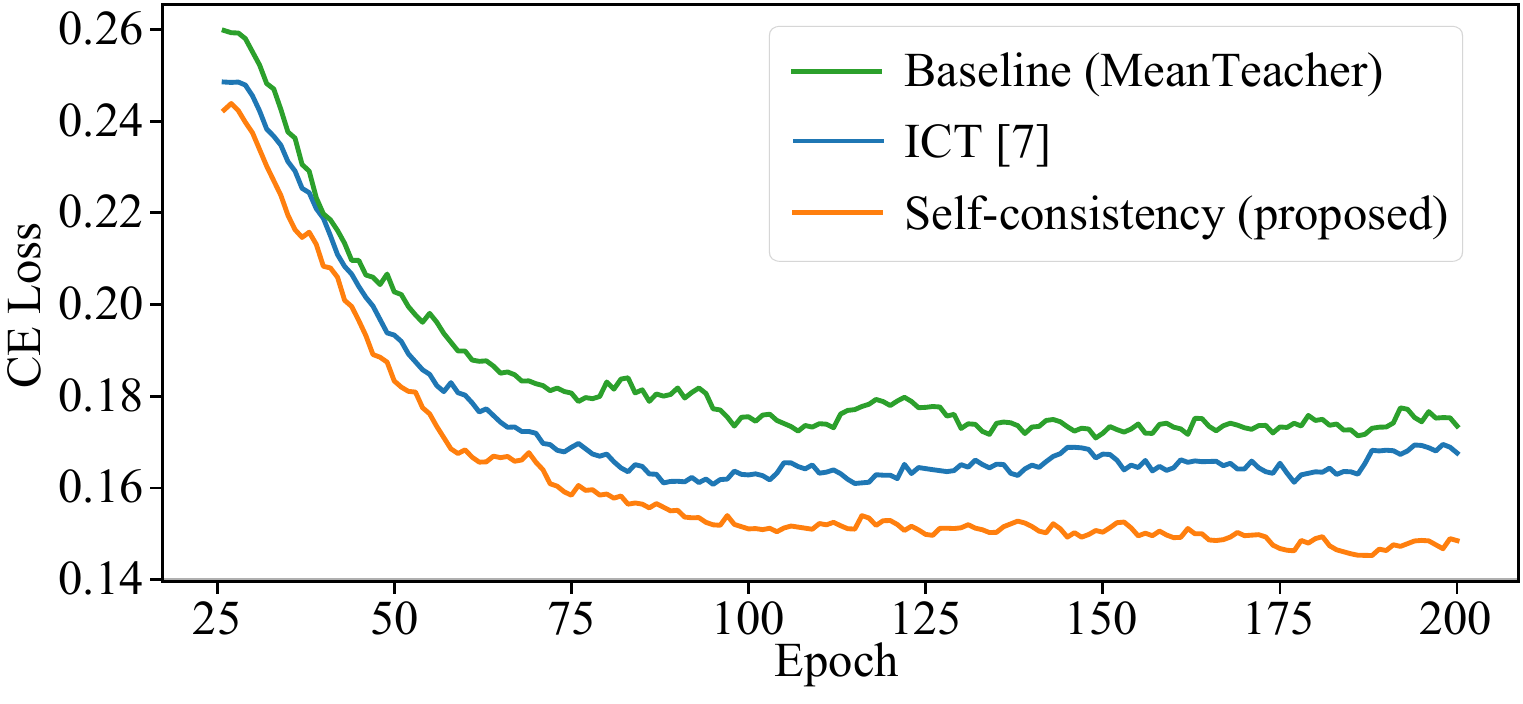}}
  \vspace{-2mm}
\end{minipage}
\caption{Cross-entropy loss of strongly-supervised validation data when training with or without self-consistency loss, comparing with the ICT scheme \cite{verma2019interpolation}.}
\label{figure: validation_loss}
\vspace{-6mm}
\end{figure}

In the random warping policy, the only hyperparameter that needs to be grid-searched is the maximum transformation magnitude $d_\text{max}$. As shown in Fig.~\ref{figure: rank}, the performance improves as the maximum transformation magnitude increases until about 5 or 6; accordingly we set $d_\text{max}$ to 5. 

Table~\ref{table: ablation_study} shows the result of ablation study, in which the proposed modules are individually added step by step. As seen, the proposed schemes including hard mixup, RandomWarping and self-consistency, lead to noticeable positive contributions. To further investigate the efficacy of the proposed methods, we also conduct experiments to replace hard mixup and self-consistency by vanilla mixup \cite{zhang2017mixup} and ICT-like consistency \cite{verma2019interpolation}, respectively. While vanilla mixup is slightly better in cross-trigger ($\text{PSDS}_2$), hard mixup gives more significant gain in response time ($\text{PSDS}_1$). The proposed self-consistency outperforms the ICT-like consistency in both metrics.

\subsection{Comparison with other semi-supervised strategies}
To evaluate the proposed SSL strategy collectively, we conduct a thorough comparison with other widely used SSL strategies, including ICT \cite{verma2019interpolation}, SCT \cite{koh2021sound} and their combination, employing the same baseline network used for the proposed strategy. Table~\ref{table: comparision} shows the comparison results, and Fig.~\ref{figure: validation_loss} gives the validation loss curves of the models with baseline MeanTeacher, ICT and self-consistency. Compared with the baseline MeanTeacher model, ICT largely improves the performance by its proposed teacher-student consistency loss, while the always-hold self-consistency loss further improves the performance, which can be reflected by a lower validation loss curve shown in Fig.~\ref{figure: validation_loss}. SCT performance is not as good as ICT, which indicates that the time shift is not as efficient as interpolation (mixup). Combining ICT and SCT does not outperform ICT alone, which implies that the naive addition of ICT and SCT \cite{koh2021sound} is not an effective way to combine multiple different augmentations schemes. In contrast, as shown in Table~\ref{table: ablation_study}, the proposed strategy is able to efficiently combine multiple different augmentations. Overall, the proposed method remarkably outperforms ICT and SCT, due to the strength of each proposed module and the efficient stochastic combination of them. 

\subsection{Comparison with DCASE2021 submissions}
To further assess the efficacy of the proposed method and conduct fair comparisons with heavily processed DCASE 2021 submitted models, we also took advantage of some existing post-processing and ensembling techniques in our model. A temperature factor of 2.1 is used for inference temperature tuning as in \cite{Zheng2021}. Median filter is applied to smooth the frame-level predictions. Following \cite{Liu2020}, the length of each median filter is calculated by $\text{length}_\text{class} = 0.55 \times \text{avg\_duration}_{\text{class}}/\text{duration}_{\text{frame}}$ and manually search for the best value with a range of $\pm 2$. The length of class-wise median filters are set to $\{3, 28, 7, 4, 7, 22, 48, 19, 10, 50\}$, for the event classes of \emph{\{alarm bell ringing, blender, cat, dishes, dog, electric shaver toothbrush, frying, running water, speech, vacuum cleaner\}}, respectively. Moreover, model ensembling was applied to fuse the predictions of multiple differently trained models. We trained eleven models with different variants of RCT: substituting time masking with frequency masking \cite{park2019specaugment}; adding FilterAug \cite{Nam2021} into audio warping choices; randomly selecting one or two methods in audio warping; and, reducing the weight of the MeanTeacher loss. We found that all different variants achieve reasonable performances. This demonstrates the flexibility of RCT and its capacity in incorporating new audio transformations into the framework with a low tuning cost overhead.

The proposed system is compared with DCASE2021 top-ranked submissions in Table~\ref{table: match_comparison}. The scores of DCASE2021 submissions are directly quoted from the challenge results. The proposed system noticeably outperforms all other systems employing the baseline CRNN network, which verifies the superiority of the proposed RCT strategy. Furthermore, as seen the performance of the proposed system is very close to the two first-ranked submissions \cite{Zheng2021, Kim2021}. They both use more powerful networks, i.e.~SKUnit and RCRNN, which were able to largely improve the performance \cite{Zheng2021, Kim2021}. The proposed framework is independent of the network and can be easily applied along with more advanced architectures to achieve higher performance.

\begin{table}[]
    \centering
    \caption{Comparing the proposed SSL strategy with other alternatives. Each score is obtained by averaging three trials.}
    \begin{tabular}{lcccc}
        \toprule
        \textbf{Model}  & \textbf{$\text{PSDS}_1$} (\%)  & \textbf{$\text{PSDS}_2$} (\%) \\
        \midrule
        Baseline \cite{jiakai2018mean}                 
        & 34.7  & 53.7 \\
        SCT \cite{koh2021sound}                         
        & 36.0  & 55.6 \\
        ICT \cite{verma2019interpolation}               
        & 37.7  & 57.7 \\
        ICT+SCT \cite{koh2021sound}                     
        & 37.0	& 58.7 \\
        \textbf{RCT} (proposed)                         
        & 40.1  & 61.4 \\
        \bottomrule
    \end{tabular}
    \label{table: comparision}
\end{table}
\begin{table}[]
    \centering
    \caption{Comparing the proposed system with DCASE2021 top-ranked submissions. All models are named in the form of network architecture plus the SSL strategy.}

    \begin{tabular}{lcccc}
        \toprule
        \textbf{Model}  & \textbf{$\text{PSDS}_1$} (\%)  & \textbf{$\text{PSDS}_2$} (\%) \\
        \midrule
        CRNN (baseline) \cite{jiakai2018mean}           & 34.7 & 53.7 \\
        FBCRNN+MLFL \cite{Tian2021}                     & 40.1 & 59.7  \\
        CRNN+IPL \cite{Gong2021}                        & 40.7 & 65.3  \\
        CRNN+DA \cite{Lu2021}                           & 41.9 & 63.8  \\
        CRNN+HeavyAug. \cite{Nam2021}                   & 43.4 & 63.9  \\
        RCRNN+NS \cite{Kim2021}                         & 45.1 & 67.9  \\
        SKUnit+ICT/SCT \cite{zhang2017mixup}            & 45.4 & 67.1  \\
        CRNN+\textbf{RCT} (proposed)                    & 44.0 & 67.1  \\
        \bottomrule
    \end{tabular}
    \label{table: match_comparison}
\end{table}

\section{Conclusion}

In this paper, we developed a novel semi-supervised learning (SSL) strategy, named random consistency training (RCT), for sound event detection (SED) task. The proposed method improves several core modules of SSL, including unsupervised training loss and data augmentation schemes. It leads to achieving competitive performance on the DCASE 2021 challenge dataset. RCT provides a generic framework which can be effectively employed along with more advanced augmentation schemes and architectures. Besides, since RCT is not task-specific, it can potentially be applied in various audio and image processing tasks which is another broad avenue for future work.

\vfill\pagebreak

  \clearpage

\bibliographystyle{IEEEtran}
\bibliography{mybib}

% \begin{thebibliography}{9}
% \bibitem[1]{Davis80-COP}
%   S.\ B.\ Davis and P.\ Mermelstein,
%   ``Comparison of parametric representation for monosyllabic word recognition in continuously spoken sentences,''
%   \textit{IEEE Transactions on Acoustics, Speech and Signal Processing}, vol.~28, no.~4, pp.~357--366, 1980.
% \bibitem[2]{Rabiner89-ATO}
%   L.\ R.\ Rabiner,
%   ``A tutorial on hidden Markov models and selected applications in speech recognition,''
%   \textit{Proceedings of the IEEE}, vol.~77, no.~2, pp.~257-286, 1989.
% \bibitem[3]{Hastie09-TEO}
%   T.\ Hastie, R.\ Tibshirani, and J.\ Friedman,
%   \textit{The Elements of Statistical Learning -- Data Mining, Inference, and Prediction}.
%   New York: Springer, 2009.
% \bibitem[4]{YourName17-XXX}
%   F.\ Lastname1, F.\ Lastname2, and F.\ Lastname3,
%   ``Title of your INTERSPEECH 2022 publication,''
%   in \textit{Interspeech 2022 -- 23\textsuperscript{rd} Annual Conference of the International Speech Communication Association, September 18-22, Incheon, Korea, Proceedings, Proceedings}, 2022, pp.~100--104.
% \end{thebibliography}

\end{document}